\documentstyle[art11]{article}
\advance \topmargin by -\headheight
\advance \topmargin by -\headsep

\evensidemargin \oddsidemargin
\begin{document}
\def\DB{{Darboux-B\"{a}cklund}~}
\def\BH{{Burgers-Hopf}~}
\def\rf#1{(\ref{eq:#1})}
\def\lab#1{\label{eq:#1}}
\def\nonu{\nonumber}
\def\br{\begin{eqnarray}}
\def\er{\end{eqnarray}}
\def\be{\begin{equation}}
\def\ee{\end{equation}}
\def\eq{\!\!\!\! &=& \!\!\!\! }
\def\foot#1{\footnotemark\footnotetext{#1}}
\def\lb{\lbrack}
\def\rb{\rbrack}
\def\llangle{\left\langle}
\def\rrangle{\right\rangle}
\def\blangle{\Bigl\langle}
\def\brangle{\Bigr\rangle}
\def\llb{\left\lbrack}
\def\rrb{\right\rbrack}
\def\lcurl{\left\{}
\def\rcurl{\right\}}
\def\({\left(}
\def\){\right)}
\newcommand{\nit}{\noindent}
\newcommand{\ct}[1]{\cite{#1}}
\newcommand{\bi}[1]{\bibitem{#1}}
\def\lskip{\vskip\baselineskip\vskip-\parskip\noindent}
\relax
\def\mskp{\par\vskip 0.3cm \par\noindent}
\def\sskp{\par\vskip 0.15cm \par\noindent}
%                     common physics symbols
\def\tr{\mathop{\rm tr}}
\def\Tr{\mathop{\rm Tr}}
\def\v{\vert}
\def\bv{\bigm\vert}
\def\Bgv{\;\Bigg\vert}
\def\bgv{\bigg\vert}
\newcommand\partder[2]{{{\partial {#1}}\over{\partial {#2}}}}
\newcommand\sbr[2]{\left\lbrack\,{#1}\, ,\,{#2}\,\right\rbrack}
\newcommand\pbr[2]{\{\,{#1}\, ,\,{#2}\,\}}
\newcommand\pbbr[2]{\lcurl\,{#1}\, ,\,{#2}\,\rcurl}
%
%                    math symbols
\def\a{\alpha}
\def\b{\beta}
\def\d{\delta}
\def\D{\Delta}
\def\eps{\epsilon}
\def\vareps{\varepsilon}
\def\g{\gamma}
\def\G{\Gamma}
\def\grad{\nabla}
\def\h{{1\over 2}}
\def\l{\lambda}
\def\L{\Lambda}
\def\m{\mu}
\def\n{\nu}
\def\o{\over}
\def\om{\omega}
\def\O{\Omega}
\def\p{\phi}
\def\P{\Phi}
\def\pa{\partial}
\def\pr{\prime}
\def\ra{\rightarrow}
\def\s{\sigma}
\def\S{\Sigma}
\def\t{\tau}
\def\th{\theta}
\def\Th{\Theta}
\def\ti{\tilde}
\def\wti{\widetilde}
\def\ca{{\cal A}}
\def\cb{{\cal B}}
\def\ce{{\cal E}}
\def\cd{{\cal D}}
\def\cH{{\cal H}}
\def\cL{{\cal L}}
%       fake blackboard bold macros for reals, complex, etc.
\def\rlx{\relax\leavevmode}
\def\inbar{\vrule height1.5ex width.4pt depth0pt}
\def\IZ{\rlx\hbox{\sf Z\kern-.4em Z}}
\def\IR{\rlx\hbox{\rm I\kern-.18em R}}
\def\IC{\rlx\hbox{\,$\inbar\kern-.3em{\rm C}$}}
\newcommand\sumi[1]{\sum_{#1}^{\infty}}   %% summation till infinity
%
%               This defines remark, proposition etc.
\def\mark{\noindent{\bf Remark.}\quad}
\def\prop{\noindent{\bf Proposition.}\quad}
\def\theor{\noindent{\bf Theorem.}\quad}
\def\name{\noindent{\bf Definition.}\quad}
\def\exam{\noindent{\bf Example.}\quad}
\def\proof{\noindent{\bf Proof.}\quad}
\def\lemma{\noindent{\bf Lemma.}\quad}
%%
%
%       This defines the journal citations
%
\newcommand\PRL[3]{{\sl Phys. Rev. Lett.} {\bf#1} (#2) #3}
\newcommand\NPB[3]{{\sl Nucl. Phys.} {\bf B#1} (#2) #3}
\newcommand\NPBFS[4]{{\sl Nucl. Phys.} {\bf B#2} [FS#1] (#3) #4}
\newcommand\CMP[3]{{\sl Commun. Math. Phys.} {\bf #1} (#2) #3}
\newcommand\PRD[3]{{\sl Phys. Rev.} {\bf D#1} (#2) #3}
\newcommand\PLA[3]{{\sl Phys. Lett.} {\bf #1A} (#2) #3}
\newcommand\PLB[3]{{\sl Phys. Lett.} {\bf #1B} (#2) #3}
\newcommand\JMP[3]{{\sl J. Math. Phys.} {\bf #1} (#2) #3}
\newcommand\PTP[3]{{\sl Prog. Theor. Phys.} {\bf #1} (#2) #3}
\newcommand\SPTP[3]{{\sl Suppl. Prog. Theor. Phys.} {\bf #1} (#2) #3}
\newcommand\AoP[3]{{\sl Ann. of Phys.} {\bf #1} (#2) #3}
\newcommand\PNAS[3]{{\sl Proc. Natl. Acad. Sci. USA} {\bf #1} (#2) #3}
\newcommand\RMP[3]{{\sl Rev. Mod. Phys.} {\bf #1} (#2) #3}
\newcommand\PR[3]{{\sl Phys. Reports} {\bf #1} (#2) #3}
\newcommand\AoM[3]{{\sl Ann. of Math.} {\bf #1} (#2) #3}
\newcommand\UMN[3]{{\sl Usp. Mat. Nauk} {\bf #1} (#2) #3}
\newcommand\FAP[3]{{\sl Funkt. Anal. Prilozheniya} {\bf #1} (#2) #3}
\newcommand\FAaIA[3]{{\sl Functional Analysis and Its Application} {\bf #1}
(#2) #3}
\newcommand\BAMS[3]{{\sl Bull. Am. Math. Soc.} {\bf #1} (#2) #3}
\newcommand\TAMS[3]{{\sl Trans. Am. Math. Soc.} {\bf #1} (#2) #3}
\newcommand\InvM[3]{{\sl Invent. Math.} {\bf #1} (#2) #3}
\newcommand\LMP[3]{{\sl Letters in Math. Phys.} {\bf #1} (#2) #3}
\newcommand\IJMPA[3]{{\sl Int. J. Mod. Phys.} {\bf A#1} (#2) #3}
\newcommand\AdM[3]{{\sl Advances in Math.} {\bf #1} (#2) #3}
\newcommand\RMaP[3]{{\sl Reports on Math. Phys.} {\bf #1} (#2) #3}
\newcommand\IJM[3]{{\sl Ill. J. Math.} {\bf #1} (#2) #3}
\newcommand\APP[3]{{\sl Acta Phys. Polon.} {\bf #1} (#2) #3}
\newcommand\TMP[3]{{\sl Theor. Mat. Phys.} {\bf #1} (#2) #3}
\newcommand\JPA[3]{{\sl J. Physics} {\bf A#1} (#2) #3}
\newcommand\JSM[3]{{\sl J. Soviet Math.} {\bf #1} (#2) #3}
\newcommand\MPLA[3]{{\sl Mod. Phys. Lett.} {\bf A#1} (#2) #3}
\newcommand\JETP[3]{{\sl Sov. Phys. JETP} {\bf #1} (#2) #3}
\newcommand\JETPL[3]{{\sl  Sov. Phys. JETP Lett.} {\bf #1} (#2) #3}
\newcommand\PHSA[3]{{\sl Physica} {\bf A#1} (#2) #3}
\newcommand\PHSD[3]{{\sl Physica} {\bf D#1} (#2) #3}
\newcommand\JPSJ[3]{{\sl J. Phys. Soc. Jpn.} {\bf #1} (#2) #3}

\vspace*{-1cm}
\noindent
October, 1995 \hfill{UICHEP-TH/95-11}\\
${}$ \hfill{solv-int/9510009} \\
\begin{center}
{\large {\bf On Integrable Models and their Interrelations}}
\end{center}
\vskip .3in
\begin{center}
{ H. Aratyn\footnotemark
\footnotetext{Talk given at the Theoretical Physics Symposium in honor of
Paulo Leal Ferreira (S\~{a}o Paulo, August 7-11,1995)}}
\par \vskip .1in \noindent
Department of Physics \\
University of Illinois at Chicago\\
845 W. Taylor St.\\
Chicago, IL 60607-7059, {\em e-mail}:
aratyn@uic.edu \\
\par \vskip .3in
E. Nissimov${}^{a),\, 2}$  and S. Pacheva${}^{a),b),}$\foot{Supported in part
by Bulgarian NSF grant {\em Ph-401}}
\par \vskip .1in \noindent
${}^{a)}$ Institute of Nuclear Research and Nuclear Energy \\   %===
Boul. Tsarigradsko Chausee 72, BG-1784 ~Sofia, Bulgaria \\
{\em e-mail}: emil@bgearn.bitnet, svetlana@bgearn.bitnet \\
and \\
${}^{b)}$ Department of Physics, Ben-Gurion University of the Negev \\  %===
Box 653, IL-84105 $\;$Beer Sheva, Israel \\
{\em e-mail}: emil@bgumail.bgu.ac.il, svetlana@bgumail.bgu.ac.il
\end{center}
\vskip .3in

\begin{abstract}
We present an elementary discussion of the Calogero-Moser
model.
This gives us an opportunity to illustrate basic concepts of the dynamical
integrable models.
Some ideas are also presented regarding interconnections
between integrable models based on the relation established
between the Calogero-Moser model and the truncated KP hierarchy of
Burgers-Hopf type.
\end{abstract}

\noindent
{\large {\bf 1. Introduction. Calogero-Moser Model}}
\mskp
The main purpose of this talk is to describe, in an elementary way,
the notion of integrability, its Lax formulation
and the relations, which can be established between the various
integrable models despite their different appearances and origins.

The subject of integrability is currently of widespread interest
in view of the recent developments in high energy physics, which brought
integrable hierarchies, including the Kadomtsev-Petviashvili one
($KP$), into the central position
in the studies of the matrix models known to describe, at multicritical points,
$c \leq 1\,$  matter systems coupled to $D=2$ quantum gravity \ct{2d}.
It is furthermore of interest to study interrelations between integrable models
as they can reveal new classes of solutions.

We first recall the notion of integrability.
\mskp
{\sl Complete integrability:}
Consider a Hamiltonian system with coordinates
$\( {\vec q}, {\vec p}\) = \( q_1, \ldots , q_N, p_1, \ldots , p_N \)$
possessing a
standard Hamiltonian structure with a Hamiltonian $ H(p,q)$ and Poisson
bracket $\{ \cdot,\cdot \}$.
We call a system {\sl integrable} if we can %find the action-angle variables
%%and
write the general
solution to the equations of motion in terms of (finitely many) algebraic
manipulations, which can include evaluation of integrals in terms of known
functions.

A condition for integrability is existence of sufficiently many, independent,
Poisson commuting (i.e. in involutions) functions  $h_r (p,q)$
($r=1, \ldots,N$) of ${\vec q}$ and ${\vec p}$, which are integrals of motion.
The fact that the functions $h_r, h_s$ are in involution:
\be
\pbr{h_r}{h_s} = \sum_{j=1}^{N} \( \partder{h_r}{q_j} \partder{h_s}{p_j}
-\partder{h_r}{p_j} \partder{h_s}{q_j} \) \, = \, 0
\lab{invo}
\ee
implies that the function $h_r$ is constant along the solutions of the
Hamiltonian system generated by  $h_s$ :
\be
{\dot q_j } =  \partder{h_s}{p_j} \qquad ;\qquad
{\dot p_j } = - \partder{h_s}{q_j}
\lab{hameq}
\ee
as follows by inserting \rf{hameq} into \rf{invo}
\be
\sum_{j=1}^{N} \( \partder{h_r}{q_j} {\dot q_j }
+\partder{h_r}{p_j}  {\dot p_j } \) = {\dot h_r} \, = \, 0
\lab{consth}
\ee

We will now illustrate the concept of integrability on the specific example
of Calogero-Moser model \ct{perel,hoppe} and show how the Lax formulation
arises
naturally in this context.
We first recall the formulation of the problem.  Let ${ X} (t)$ be a
$N \times N$ Hermitian matrix. We define a model by choosing a most simple
dynamics for the matrix ${ X} (t)$. Namely, we let the second time derivative
of $X$ be zero:
\be
{\ddot { X}} (t) =0 \quad \to \quad { X} (t) = { X} (0) +
{\dot { X}} (0)\, t
\lab{ddot}
\ee
The dynamics will be described in terms of the eigenvalues
of the ${ X}$-matrix. The next step involves, therefore, a diagonalization
of ${ X} (t)$ by a unitary matrix $U$.
\be
{ X} \quad \to \quad { Q}(t) = U^{-1} (t) X (t) U (t)
= \left(\begin{array}{ccc}
q_1 (t)  &  & \\
  & \ddots &  \\
  &  & q_N (t)
\end{array} \right)
\lab{uxu}
\ee
where we denoted by $\{ q_j \; \v j=1, \ldots , N\}$ the eigenvalues
of the $X$ matrix. Making use of the identity
$\pa U^{-1} (t) / \pa t = - U^{-1} (t) (\pa U (t) / \pa t) U^{-1} (t)$
we find the flows of $X$:
\be
\partder{ X}{t} = \sbr{(\pa U (t) / \pa t) U^{-1} (t)}{U (t) Q U^{-1} }
+ U (t)\partder{ Q}{t} U^{-1} (t)= U (t) L ({t}) U^{-1} (t)
\lab{xovert}
\ee
where we have introduced the matrix $L$, which is a prototype of a Lax
operator:
\be
L \equiv \sbr{U^{-1} (t) (\pa U (t) / \pa t)}{Q (t) }
+ \partder{ Q}{t} = \sbr{M (t)}{Q (t) }
+ \partder{ Q}{t}
\lab{lmat}
\ee
with $ M (t)\equiv  U^{-1} (t) (\pa U (t) / \pa t)$.
Differentiating \rf{xovert} one more time we obtain:
\be
0= {\ddot X} = U \( \sbr{U^{-1} (t) (\pa U (t) / \pa t)}{L (t) } +
{{\dot L}} \) U^{-1}
\lab{xddot}
\ee
which  implies the Lax equation of motion:
\be
{\dot L} =\sbr{L (t) }{M (t)}
\lab{laxeq}
\ee
The fact that we can cast the flows of dynamical variables of the integrable
model in the form of Lax equation signals integrability.
In fact, it can be shown that any completely integrable Hamiltonian system
admits a Lax representation (at least locally) \ct{BV90}.
The Lax formulation leads straightforwardly to the construction of the
integrals of motion. Namely, for any invariant function $I$, like
$I(A) = \Tr (A^k)$ for some matrix $A$, $I \( L\) $ is a constant of motion.

For simplicity, we assume that $U (t=0) =1$, which defines as initial
conditions: $Q(0)= X (0)$ and $L(0)= {\dot X} (0)$.
Since $\ddot{X}=0$ the matrix $C = \sbr{X}{\dot{X}}$ is a constant
and therefore given by the initial conditions:
\be
C_{ij} = \(\sbr{Q(0)}{L(0)}\)_{ij} \quad \to \quad
{L(0)}_{ij} = { C_{ij} \over q_i (0) - q_j  (0) } \quad i \ne j
\lab{lzeroij}
\ee
{}From \rf{lmat} we find:
\be
{L(t)}_{ij} = \d_{ij} {\dot q}_j - M_{ij}(t) \(q_i (t) - q_j  (t) \)
\quad \to \quad M_{ij}(0) =- { C_{ij} \over \( q_i (0) - q_j  (0)\)^2 }\quad
\;i \ne j
\lab{mijzero}
\ee
Extending this straightforwardly to arbitrary time $t$ we find
\be
M_{ij}(t) =- { C_{ij} \over \( q_i (t) - q_j  (t)\)^2 }\quad;\quad i \ne j
\lab{mijt}
\ee
which provides an ansatz consistent with the off-diagonal part of \rf{laxeq},
when verified together with $C_{ij} = i g (1 - \d_{ij})$ and
$M_{ii} = ig \sum_{k \ne j} \( q_j (t) - q_k  (t)\)^{-2}$
($g$ is a coupling constant).
With these assumptions we construct the Lax pair
\br
{L}_{ij} (t)\eq \d_{ij} {\dot q}_j +{ i g (1 - \d_{ij})\over
\(q_i (t) - q_j  (t) \)} \lab{lijt}\\
{M}_{ij} (t)\eq ig \d_{ij} \sum_{k \ne i} \( q_i (t) - q_k  (t)\)^{-2}
- { i g (1 - \d_{ij}) \over \( q_i (t) - q_j  (t)\)^2 }
\lab{mijta}
\er
which, when plugged in \rf{laxeq}, produces equations of motion.
These  equations of motion can alternatively be produced by inserting the
Hamiltonian:
\be
\cH = \h \( \sum_{i=1}^N p_i^2 + g^2 \sum_{i \ne j}
{ 1 \over \( q_i (t) - q_j  (t)\)^2 } \) = \h \Tr ( L^2)
\lab{hamfct}
\ee
into the Hamilton equations of motion:
\be
{\dot p_i} = \pbr{\cH}{p_i} = 2 \sum_{i \ne j} { g^2
 \over \( q_i (t) - q_j  (t)\)^3 } \quad ; \quad
{\dot q_i} = \pbr{\cH}{q_i} = p_i
\lab{hameqs}
\ee
The system is completely integrable in a sense that the above equations of
motion can be solved by eigenvalues of ${ X} (t) = { Q} (0) +L (0)\, t$
with  $ { Q}_{ij}  (0) = \d_{ij}  q_j (0) $ and $ { L}_{ij}  (t=0)$ given by
\rf{lijt}.

Having established the Lax representation for the equations of motion it is
easy to find the integrals of motion following the standard recipe:
\be
H_k =   \Tr ( L^k)  \quad \to \quad \partder{H_k}{t}
= k  \Tr ( \sbr{L}{M} L^{k-1}) = 0
\lab{recipe}
\ee
The presence of higher Hamiltonians signals that there exist
``higher'' times $t_k$. The corresponding higher
flows are also governed by the Lax equations:
\be
\partder{L}{t_k} =\sbr{L (t) }{M_k (t)}
\lab{laxeqe}
\ee
with
\be
(M_k)_{ij} =  \d_{ij} \sum_{k \ne i} {k \over  q_i (t) - q_k  (t)}
\(L^{k-1}\)_{ik} - (1 - \d_{ij}){k \over  q_i (t) - q_j  (t)} \(L^{k-1}\)_{ij}
\lab{mijte}
\ee
The Lax equations \rf{laxeqe} allow now a direct proof that $H_k$'s are in
involution:
\be
\pbr{H_k}{H_l} =   \pbr{H_k}{\Tr ( L^l)}= \partder{\Tr ( L^l)}{t_k}
=  \Tr ( \sbr{L^l}{M_k} ) = 0
\lab{recip}
\ee
To investigate further the algebraic structure of the Calogero-Moser model
we define in addition to $H_k$ also  $\cL_k = \Tr ( Q L^{k+1}) $
for $k=-1,0, \ldots$, where $Q$ is the diagonal matrix defined in \rf{uxu}.
It can be shown \ct{wadati} that $H_k$ and $\cL_k $ enter the Poisson algebra
identical to that of the algebra
being a semi-direct product of an Abelian Kac-Moody current algebra with the
(centerless) Virasoro algebra:
\be
\pbr{H_n}{\cL_m} = n H_{n+m} \quad; \quad \pbr{\cL_n}{\cL_m}=
(n-m) \cL_{n+m}
\lab{kmvalg}
\ee
These results follows from the form of flows for $L$ \rf{laxeqe} and
for $Q$ \ct{regge,wadati}:
\be
\partder{Q}{t_n} =\sbr{Q }{M_n} + n L^{n-1}
\lab{qeqe}
\ee
Similar structure has also been found in the hierarchy of nonlinear
partial differential equations which serves as our second main illustration.
\mskp
{\large {\bf 2. The ${\bf KP}$ hierarchy}}
\mskp
An important example of an integrable system admitting
the Lax formulation is given by the $KP$ hierarchy consisting of
the pseudo-differential Lax  operator $L$ :
\be
L = D + \sum_{i = 1}^{\infty} u_i D^{-i} \; .
\lab{kpl}
\ee
which enters the following family of Lax equations:
\be
\pa_n L = \partder{L}{t_n} = \lb B_n \, , \, L \rb
\quad    \; \; n = 1, 2, \ldots \lab{lax-eq}
\ee
describing isospectral deformations of $L$.
In \rf{lax-eq} $t = \{ t_n \} $ are the evolution parameters
(infinitely many time coordinates) and
$B_n = L^n_{+} $ is the differential part of
$L^n = L^n_{+} + L^n_{-} = \sumi{i=0} P_i (n) D^i +
\sum_{-\infty}^{-1} P_i (n) D^i$.

One can also represent the Lax operator in terms of the dressing
operator $W= 1 + \sum_1^{\infty} w_n D^{-n}$ through
$L = W D\,W^{-1}$.
In this framework the equation \rf{lax-eq} is equivalent to the so called
Wilson-Sato equation:
\be
\pa_n W \equiv \frac{\pa W}{\pa t_n} = B_n W - W D^n
= - L^n_{-} W
\lab{sato-a}
\ee
Define next the Baker-Akhiezer (BA) function via
\be
\psi (t,\lambda) = W\,e^{\xi} = w(t,\lambda)e^{\xi};\, \ \,
\, \ \,w(t,\lambda) = 1 + \sum_1^{\infty}w_n(t)\lambda^{-n} \ ,
\lab{BA}
\ee
where
\be
\xi(t,\lambda) \equiv  \sum_{n=1}^\infty t_n\lambda^n \qquad; \quad t_1 = x
\lab{xidef}
\ee
There is also an adjoint wave function
$\psi^{*} = W^{*-1} \exp(-\xi (t,\l )) =
w^{*}(t,\lambda)\exp(-\xi (t,\l )),\,\,w^{*}(t,\lambda) = 1 +
\sum_1^{\infty}w_i^{*}(t)\lambda^{-i}$,
and one has the following linear systems:
\be
L\psi =
\lambda\psi;\, \ \,\partial_n\psi = B_n\psi;\, \ \,L^{*}\psi^{*} =
\lambda\psi^{*};\, \ \,\partial_n\psi^{*} = -B_n^{*}\psi^{*} \ .
\lab{linsys}
\ee
Note that eq.\rf{lax-eq} for the KP hierarchy flows
follows then from the compatibility conditions among these equations.

Also, the KP hierarchy has an ``additional" symmetry structure
\ct{Orlovetal,moerbeke,dickey} similar to the one encountered in the
Calogero-Moser model.
Define namely
\be
Q \equiv  W \( \sum k t_k D^{k-1} \) W^{-1}
\lab{kpq}
\ee
We can now supplement \rf{linsys} by
\be
Q \psi = \partder{\psi}{\l} \quad \to \quad \sbr{L}{Q}=1
\lab{qpsi}
\ee
$Q$ enters the evolution equations
\be
\partder{Q}{t_n} = \sbr{B_n}{Q } =\sbr{Q }{(L^n)_{-}} + n L^{n-1}
\lab{kpqeqe}
\ee
which have the same form as \rf{qeqe} in the setting of the
Calogero-Moser model.

There exists a quite natural way
of describing the KP hierarchy based on one single function
-- the so-called tau function $\tau(t)$.
This approach is an alternative to using the Lax operator
and the calculus of pseudo-differential operators.
The $\t$-function is related to the BA functions via
\br
\psi(t,\l) \eq
e^{\xi(t,\l)} { \tau \(t_i - 1/i \l^{i}\)\over \tau (t_i)}
= e^{\xi(t,\l)} \sumi{n=0} { p_n \( - {\ti \pa}\)
\tau (t)\over \tau (t)} \l^{-n} \lab{psi-main} \\
\psi^{*}(t,\l) \eq
e^{\xi(t,\l)} { \tau \(t_i + 1/i \l^{i}\)\over \tau (t_i)}
= e^{\xi(t,\l)} \sumi{n=0} { p_n \(  {\ti \pa}\)
\tau (t)\over \tau (t)} \l^{-n} \lab{psi-mainc}
\er
where $\ti{\pa} = (\pa_1,(1/2)\pa_2,(1/3)\pa_3, \ldots)$
and the Schur polynomials $p_n$ are defined through \\
$ e^{\xi (t,\l )}=  \sumi{n=0} p_n (t_1, t_2, \ldots ,t_j) \l^j $.
The BA functions enter the fundamental bilinear identity
\be
\oint_{\infty}\psi(t,\l) \psi^{*}(t',\l) d\l = 0
\lab{bilide}
\ee
which generates the entire KP hierarchy.
In \rf{bilide} $\oint_{\infty}(\cdot)d\l$ is the residue integral about
$\infty$.
It is possible to rewrite the above identity in terms of the tau functions
obtaining
\be
\oint_{\infty}\tau(t-[\l^{-1}]) \tau(t'+[\l^{-1}])
e^{\xi(t,\l)-\xi(t',\l)}d\l = 0
\lab{biltau}
\ee
Taylor expanding \rf{biltau} in $y$ ($t\to t-y,\,\,t'\to t+y$)
leads to
\be
\(\sumi{0} p_n(-2y) p_{n+1} (\ti D) e^{\sum_1^{\infty} y_i D_i}\)
\tau\,\cdot\,\tau = 0
\lab{hir1}
\ee
where $D_i$ is the Hirota derivative defined as
\be
D^m_j a\,\cdot\,b = (\pa^m/\pa s_j^m) a(t_j+s_j)b(t_j-s_j)\v_{s=0}
\quad ,\quad     \ti D = (D_1,(1/2) D_2,(1/3)D_3,\ldots)
\lab{hir-der}
\ee
The coefficients of the $y_n$-expansion in \rf{hir1} yield
\be
\( \h D_1D_n -  p_{n+1} (\ti D) \) \tau \cdot \tau = 0
\lab{hir2}
\ee
which are called the Hirota bilinear equations.
\mskp
{\large {\bf 3. Truncated ${\bf KP}$ Hierarchy, Burgers-Hopf hierarchy}}
\mskp
Here we consider a class of truncated KP hierarchies constructed
from $m$-truncated dressing operator $ W $:
\be
W =  \sum_{i=0}^{m} w_i (t) D^{-i}\qquad;\qquad w_i =
{p_i (- {\ti \pa}) \tau (t)   \over \tau (t)}
\lab{dress-tr}
\ee
The Lax operator is given by the usual relation $L = W D W^{-1}$
and the $m$-truncated dressing operator $ W $ satisfies the Sato equations
as in \rf{sato-a}.

The generalized Hopf-Cole transformation applied to this problem leads
to the differential equation:
\be
W \pa^m \p_k = 0 \qquad k =1, \ldots, m
\lab{hopf-co}
\ee
It is well-known that, while solutions of the general KP hierarchy
form the universal Grassmann manifold UGM, the solutions of \rf{hopf-co}
defining the $m$-truncated KP hierarchy form the Grassmann manifold
$ GM (m, \infty ) = {\rm Mat } (\infty \times m ) / GL (m; \IC)$
where ${\rm Mat } (\infty \times m ) $ denotes $\infty \times m$
matrices of rank $m$ \ct{harada,ohta}.

In terms of the solutions of the $m$-th order differential equation
\rf{hopf-co} the Wilson-Sato equations take a simpler form:
\be
\pa_n \p_i = \pa^n \p_i \qquad i=1, \ldots, m
\lab{sato-lin}
\ee
In different words we have the following lemma:\\
{\bf Lemma.}\quad
{\em Eq.\rf{sato-lin} is equivalent to the Wilson-Sato equation
\rf{sato-a} for the $m$-truncated dressing operator from \rf{dress-tr}}.
\mskp
Let us return to \rf{hopf-co}. For $W = 1+ w_1 D^{-1} + \cdots + w_m D^{-m}$
equation \rf{hopf-co} factorizes as follows
\be
WD^m \p_m = \( D^m + w_1 D^{m-1} + \cdots + w_m \) \p_k =
\( D + v_m \) \( D + v_{m-1} \) \cdots \( D + v_1 \) \p_k = 0
\lab{hopf-coa}
\ee
There is a relation between the coefficients $v_i$ of \rf{hopf-coa}
and the solutions $\p_k$:
\be
v_i = \pa \( \ln { W_{i-1} \lb \p_1, \ldots, \p_{i-1} \rb \over
  W_{i}\lb \p_1, \ldots, \p_{i} \rb } \) \qquad W_0 =1
\lab{wil}
\ee
For $m=1$ this relation takes the form of the classical Hopf-Cole
transformation:
\be
\( 1 + w_1 D^{-1} \)D \p = 0 \quad \to \quad
\( \pa + w_1 \) \p = 0 \quad \to \quad w_1 = - \pa_x \ln \p
\lab{hopf-cole}
\ee
The corresponding differential operator takes the form
$WD^m = D -\pa_x \(\ln \p \) = \p D \p^{-1}$.
Since for all $m$ we have $WDW^{-1} = (WD^m) D (WD^m)^{-1}$ we are lead
to the Lax operator:
\be
L^{(1)} = \(\p D \p^{-1}\)  \; D \; \(\p D^{-1} {\p}^{-1}\)
\, =\, D+ \llb \p \( \ln \p\)^{\pr \pr} \rrb D^{-1} \(\p\)^{-1}
\lab{lone}
\ee
One finds from the above Lemma that the Lax equations \rf{lax-eq} for
$L^{(1)} $ are equivalent to $\pa_n \p = \pa_x^n \p$
or in terms of the coefficient $w \equiv -w_1$ of the dressing operator
($W = 1 + w_1 D^{-1}$) :
\br
\pa_n w \eq  \pa_x P_{n} ( w) \lab{wfaa} \\
P_{n+1} ( w) \eq (\pa + w) P_n ( w) \;\;\; n = 0,1,2 ,\ldots \qquad
P_0 (w) = 1
\lab{faa}
\er
Here $P_n (w)$ are Fa\'a di Bruno polynomials fully determined by the
recurrence relation in \rf{faa}.
The system of nonlinear differential equations in \rf{wfaa} is called
Burgers-Hopf hierarchy.

We now relate a class of solution of the \BH hierarchy to the Calogero-Moser
model \ct{chudnovsky}.
Let
\be
\p = \prod_{i \in I} \( x - q_i (t^{\pr}) \)
\lab{mero}
\ee
be a solution of \rf{sato-lin} with $x=t_1$ and
$t^{\pr} = (t_2, t_3, \ldots)$. Note that this is an ansatz for the
$\t$-function of the \BH hierarchy.
We find that equation \rf{sato-lin} is equivalent to following evolution
equation for $q_i$:
\be
\pa_n q_i = - n!
\sum_{\{j_1 < \ldots < j_{n-1}\}}^{ \{j_1, \ldots, j_{n-1}\} \not\ni i}
\(q_i - q_{j_1}\)^{-1} \cdots \(q_i - q_{j_{n-1}}\)^{-1}
\qquad ;\qquad n \geq 2
\lab{panqi}
\ee
For $n=2 $ we get:
\be
{\dot q_i} = \pa_2 q_i = - 2 \sum_{j \ne i} {1 \over \(q_i - q_{j}\)}
\lab{patwoqi}
\ee
which leads to
\be
{\ddot q_i} = 8 \sum_{j \ne i} {1 \over \(q_i - q_{j}\)^3}
\lab{papatwoqi}
\ee
Hence, the solution of \BH hierarchy of the type shown in \rf{mero}
can be embedded in the Hamiltonian system \rf{hameqs}.
Equation \rf{patwoqi}, when compared with
\rf{papatwoqi}, suggests to obtain the system from \rf{mero}
via hamiltonian reduction of Calogero-Moser model by imposing
 constraints $ \varphi_i =\p_i -2 \sum_{j \ne i} 1/ \(q_i - q_{j}\)=0$ .
These constraints turn out to be first class and consequently
the class of solutions  \rf{mero} of the \BH model can not be obtained
from the Calogero-Moser model by Hamiltonian reduction. In fact,
imposing $\varphi_i = 0$ puts all Hamiltonians $H_k$ of the Calogero-Moser
model to zero.

If, however, \rf{mero} is extended to be an ansatz for the $\t$-function of
the complete KP hierarchy (with untruncated dressing operator) then,
as shown in \ct{shiota}, the flows
of $q_i$'s obey the equations of motion \rf{hameqs} with Calogero-Moser
Hamiltonians given in \rf{recipe}.

We generalize now the \BH hierarchy by
applying successively the \DB transformations
to the \BH Lax structure \rf{lone}.
This leads to a special realization of the
truncated KP hierarchy in terms of one function $\p$ only.
The resulting hierarchy we call generalized \BH hierarchy.
The Lax structure obtained in this process
takes the following form of recursive relations:
\br
L^{(k+1)} \eq \(\Phi^{(k)}  D {\Phi^{(k)} }^{-1}\)  \; L^{(k)}
 \; \(\Phi^{(k)}  D^{-1} {\Phi^{(k)} }^{-1}\)
= D + \Phi^{(k+1)} D^{-1} \Psi^{(k+1)}
\lab{lkplus} \\
\Phi^{(k+1)} \eq \Phi^{(k)} \( \ln \Phi^{(k)}\)^{\pr \pr} + \(\Phi^{(k)}\)^2
\Psi^{(k)} \quad ,\quad \Psi^{(k+1)} = \(\Phi^{(k)}\)^{-1} \lab{pkplus}
\er
with $\Phi^{(0)} = \p$.

It is known \ct{us} that the \DB transformations induce semi-infinite
Toda chain structure on the sequence of the \BH hierarchies.
Consequently the $\t$-function of the semi-infinite Toda chain
belongs to the \DB group orbit of the trivial $\t$-function (vacuum) of the
generalized \BH system.

These remarks obtain a special relevance in view of the
relation of the generalized \BH hierarchy to the one-matrix string model.
We shall consider one-matrix model with partition function
($M$ is a Hermitian $N \times N$ matrix) :
\br
Z_N \lb t \rb \eq \int dM \exp \lcurl \sumi{r=1} t_r \Tr M^r \rcurl  \nonu\\
\eq { 1 \over N !} \int \prod_{i=1}^N d \l_i
 \exp \lcurl \( \sum_{i=1}^{N} \sumi{k=1} t_k \l^k\) \rcurl
\prod_{i >j =1}^N \( \l_i - \l_j \)^2
\lab{partl}
\er
which gives rise to a Toda matrix hierarchy system of evolution equations:
\be
\partder{}{t_r} Q = \llb Q^r_{(+)} , Q \rrb    \lab{L-3}
\ee
Choosing parametrization $Q_{nn} = a_0 (n)\;\;, \;\; Q_{n,n-1} = a_1 (n)$
(the rest being zero) for the matrix $Q$, we can cast
the matrix hierarchy equations \rf{L-3} in the form of a discrete linear system
\br
\pa \Psi_n \eq \Psi_{n+1} + a_0 (n) \Psi_n     \lab{speca}\\
\l \Psi_n \eq \Psi_{n+1}  + a_0 (n) \Psi_n
+ a_1 (n) \Psi_{n-1} \nonu \\
\eq L_n \Psi_{n} \qquad \qquad n   \geq 0
\lab{specb}
\er
where the Lax operator $L_n$ associated to the site $n$ can be written as
\be
L_n = \pa + a_1 (n) {1 \over \pa - a_0 (n-1)}
\lab{lnn}
\ee
and with compatibility conditions for \rf{speca}--\rf{specb}  having the form
of the Toda lattice equations of motion:
\br
\pa a_0 (n) \eq a_1 (n+1) - a_1 (n)  \lab{t1eqa} \\
\pa a_1 (n) \eq a_1 (n) \( a_0 (n) - a_0 (n-1) \)  \lab{t1eqb}
\er
In fact the lattice jump $n \to n+1$ can be given a meaning of
Darboux-B\"{a}cklund transformation \ct{us} within the generalized
\BH hierarchy.
To see this we rewrite \rf{speca} as follows:
\be
\Psi_{n+1}= e^{\int a_0 (n)} \pa\; e^{-\int a_0 (n)} \Psi_{n}
\lab{spectodc}
\ee
or in a equivalent form obtained taking into account \rf{t1eqb}
\be
\Psi_{n+1}= a_1 (n) e^{\int a_0 (n-1)} \pa \( a_1 (n)
e^{\int a_0 (n-1)}\)^{-1} \Psi_{n}
= \P (n) \pa \P^{-1} (n) \Psi_{n} \equiv T (n) \Psi_{n}
\lab{todadb}
\ee
where $\P (n) =a_1 (n) e^{\int a_0 (n-1)} $ and
$T (n)= \P (n) \pa \P^{-1} (n) $
plays a role of the Darboux-B\"{a}cklund transformation operator generating
the lattice translation $n \to n+1$.
We find
\br
L_{n+1} \Psi_{n+1} &=& \l \( \pa - a_0 (n) \) \Psi_n \nonu\\
&=& \( \pa - a_0 (n) \) L_n \Psi_n \nonu\\
&=& \( \pa - a_0 (n) \) L_n \( \pa - a_0 (n) \)^{-1} \Psi_{n+1}
\lab{lntoln1}
\er
So, the Lax operators at different sites are related by a \DB transformation:
\be
L_{n+1} = \( \pa - a_0 (n) \) L_n \( \pa - a_0 (n) \)^{-1}=
T (n) L_n T^{-1} (n)
\lab{transf3}
\ee
where $L_n$ itself takes the form:
\be
L_n  = \pa + \P (n) \pa^{-1}  \Psi (n)  \qquad , \quad
\P (n) =a_1 (n) e^{\int a_0 (n-1)}       \quad ,\quad
\Psi (n) \equiv  e^{-\int a_0 (n-1)}
\lab{specln}
\ee
Here $L_n$ has a form of the operator belonging to
the generalized \BH hierarchy.
Recalling that $a_1 (n=0)=0$ and therefore $L_0 = \pa$
we see that we proved:\\
\prop {\em The one-matrix model problem is equivalent
to the generalized \BH system possessing the symmetry with respect to
the \DB transformations.}

This identifies the partition function of the discrete one-matrix model
as a $\t$ function of the generalized \BH system having Wronskian form \ct{us}.
\mskp
This talk focused on the linkage between various integrable systems.
{}From the simple realization of integrability in the Calogero-Moser model
to the complex structure of the KP hierarchy of the non-linear differential
equations we have emphasized the common features involving the Lax pair
formulation and the $\t$-function construction.
The relations between different models like discrete matrix models, Toda
hierarchy and continuum hierarchies of non-linear differential equations
are of current interest in theoretical physics and are among the important
tools used to find the $\t$-function solutions of integrable
models.

\mskp
{\bf Acknowledgements}\\
HA acknowledges warm hospitality of Prof. P. Leal Ferreira and IFT-Unesp
enjoyed during numerous visits.
E.N. gratefully acknowledges support and hospitality by the {\em Deutscher
Akademischer Austauschdienst} and Prof. K.Pohlmeyer at the University of
Freiburg.

\end{document}